\documentstyle[aps,multicol,epsf,rotate]{revtex}
\begin{document}

\draft

\title{Dependence of Conductance on Percolation Backbone Mass}

\author{Gerald Paul,$^1$\thanks{Electronic address: gerryp@bu.edu}
Sergey V. Buldyrev,$^1$ Nikolay V. Dokholyan,$^1$\thanks{Present
address: Department of Chemistry and Chemical Biology, Harvard
University, Cambridge, MA 02138 USA} Shlomo Havlin,$^2$\\ Peter
R. King,$^{3,4}$ Youngki Lee,$^1$ and H. Eugene Stanley$^1$}

\address{$^1$Center for Polymer Studies and Dept. of Physics,
Boston University, Boston, MA 02215 USA\\
$^2$Minerva Center and Department of Physics, Bar-Ilan University, Ramat
Gan, Isreal\\
$^3$BP Amoco Exploration Operating Company Ltd.,
Sunbury-on-Thames, Middx., TW16 7LN, UK\\
$^4$Department of Engineering,
Cambridge University, Cambridge, UK} 

\date{\today}

\maketitle

\begin{abstract}

On two-dimensional percolation clusters at the percolation threshold, we
study $\langle\sigma(M_B,r)\rangle$, the average conductance of the
backbone, defined by two points separated by Euclidean distance $r$, of
mass $M_B$. We find that with increasing $M_B$ and for fixed $r$,
$\langle\sigma(M_B,r)\rangle$ asymptotically {\it decreases\/} to a
constant, in contrast with the behavior of homogeneous sytems and
non-random fractals (such as the Sierpinski gasket) in which conductance
increases with increasing $M_B$. We explain this behavior by studying
the distribution of shortest paths between the two points on clusters
with a given $M_B$. We also study the dependence of conductance on $M_B$
slightly above the percolation threshold.

\end{abstract}


\begin{multicols}{2}

\section{Introduction}

There has been considerable study of the bond percolation cluster
considered as a random-resistor network, with each occupied bond having
unit resistance and non-occupied bonds having infinite resistance
\cite{Herrmann86,Stauffer94,Bunde96}.  In two dimensions, the
configuration studied is typically an $L\times L$ lattice and the
conductance is measured between two opposite sides which are assumed to
have infinite conductance \cite{Alexander82,Pandey84,Duering91,Hong84,%
Derrida82,Herrmann84,Normand88,Fogelholm80,Gingold90,Fisch78,Essam85,%
Harris84,Wang86}. The backbone of the cluster is then defined as the set
of bonds that are connected to the two sides having infinite conductance
through paths that have no common bond. 

At the percolation threshold, the backbone mass scales as $\langle
M_B\rangle\sim L^{d_B}$ with $d_B=1.6432\pm 0.0008$ \cite{Grassberger99}
and in this ``bus bar'' geometry is strongly correlated with $L$.  The
total conductance of the backbone as a function of $L$ has been studied
extensively and has been found to scale as $\langle\sigma\rangle\sim
L^{-\tilde\mu}$ with $\tilde\mu=0.9826\pm 0.0008$ \cite{Grassberger99}.

\begin{figure}

\centerline{ \epsfxsize=14.0cm
\rotate[r]{ \epsfbox{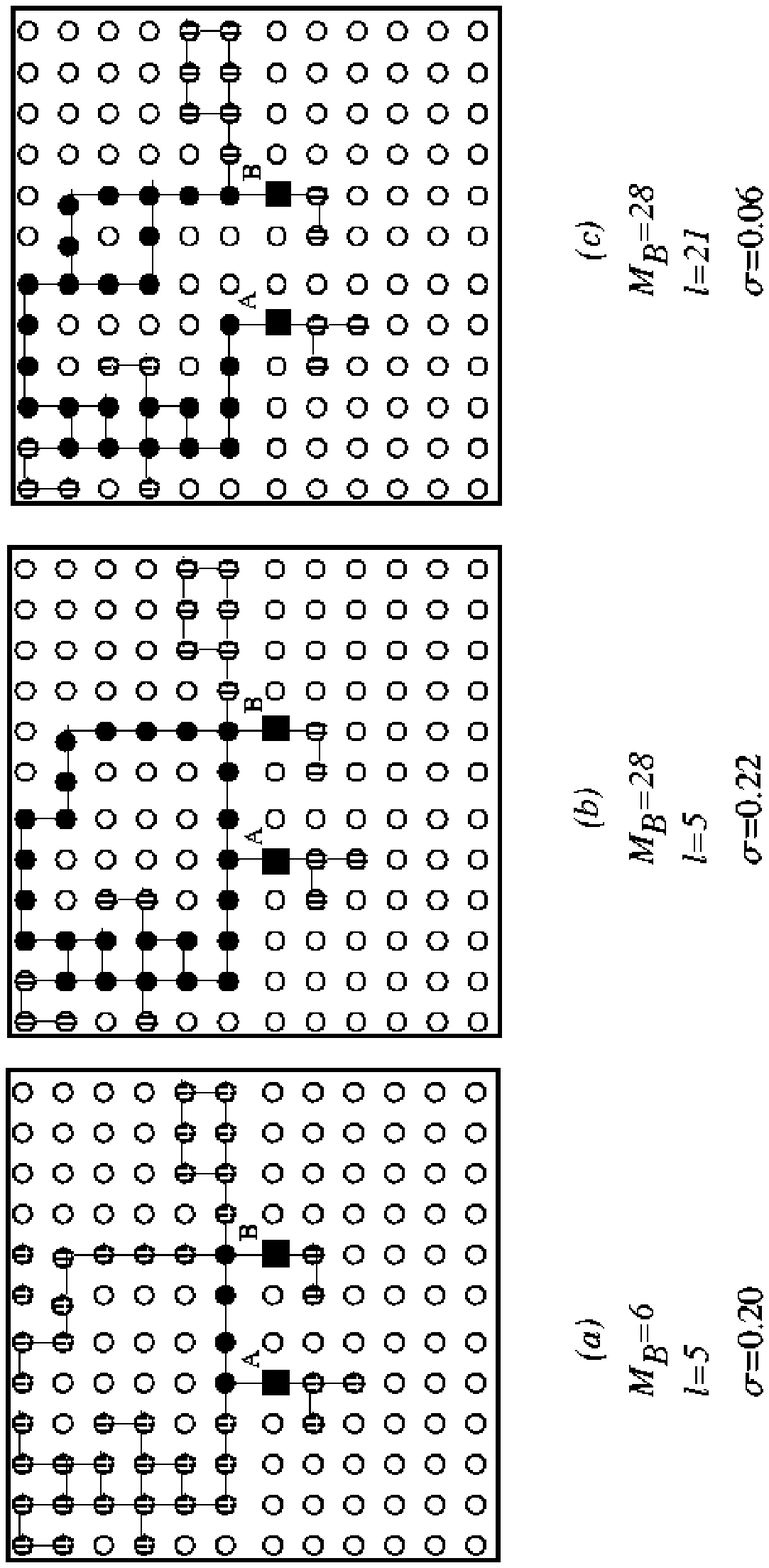} }
\vspace*{1.0cm}
}

\caption{
Typical percolation clusters. The striped sites and the black sites are
both part of the percolation cluster; only the black sites form the
cluster backbone defined by sites $A$ and $B$ (black squares). Note in
(a), with backbone size 6, the shortest path,$\ell$, between $A$ and $B$
is 5---the length of the shortest path must always be less than the
backbone mass; (b) and (c) illustrate that in clusters with large
backbones, the length of the shortest path between $A$ and $B$ can take
on a broad range of values. }
\label{fig1}
\end{figure}

Recently, the distribution of masses of backbones defined by two {\it
points}, i.e., backbones defined as the set of those bonds that are
connected by paths having no common bonds to two points separated by
distance $r$ within an $L\times L$ lattice, has been studied
\cite{Barthelemy99}. One finds that when $r\ll L$, there is a very broad
distribution of backbone masses for a given $r$.  Figure \ref{fig1}
illustrates some typical percolation clusters and their backbones
defined in this configuration.  Because of the broad distribution of
backbone masses we have the opportunity to study the conductance between
these two points separated by a fixed distance, $r$, as a function of
the mass of the backbone defined by these points.

One might expect that, for fixed $r$, the average conductance would
increase with increasing backbone mass because there could be more paths
through which current can flow. In fact, we find that the average
conductance decreases monotonically with increasing backbone size, in
contrast with the behavior of homogeneous systems and non-random
fractals in which conductance increases with increasing $M_B$. We
explain our finding by noting that the conductance is strongly
correlated with the shortest path between the two points, and then
studying the distribution of shortest paths between the two points for a
given $M_B$. This analysis extends recent studies of the distribution of
shortest paths where no restriction on $M_B$ is placed
\cite{Dokholyan98,Porto98,GrassbergerXX,ZiffXX}.

\section{Simulations}

Our system is a two-dimensional square lattice of side $L=1000$ with
points $A$ and $B$ defined as $A=(L-r/2,\ 500)$, $B=(L+r/2,\ 500)$. For
each realization of bond percolation on this lattice, if there is a path
of connected bonds between $A$ and $B$, we calculate (i) the length of
the shortest path between $A$ and $B$, (ii) the size of the backbone
defined by $A$ and $B$ and (iii) the total conductance between $A$ and
$B$. We obtain data from 100,000 realizations for each of 8 values of
$r$ (1, 2, 4, 8, 16, 32, 64, and 128) at the percolation threshold,
$p_c=0.5$. We bin these results based on the value of the backbone mass,
$M_B$ by combining results for all realizations with $2^n<M_B<2^{n+1}$
and taking the center of each bin as the value of $M_B$.

In Fig.~\ref{fig2}(a), we plot the simulation results for the average
conductance $\langle\sigma(M_B,r)\rangle$ and find that the conductance,
in fact, decreases with increasing $M_B$. The decrease is seen more
clearly in Fig.~\ref{fig2}(b), in which we plot scaled values as
discussed below.

\section{Sierpinski Gasket}

In non-fractal systems, the conductance increases as the mass of the
conductor increases. We next consider the average conductance on the
Sierpinski gasket, a non-random fractal, the first three generations of
which are illustrated in Fig.~\ref{fig3}(a)--(c).  Because the
Sierpinksi gasket is not translationally invariant, the analogue of the
average conductivity between two points in the percolation cluster is
the conductivity averaged over all pairs of points separated by distance
$r$ . At each successive generation, there are two types of pairs: (i)
pairs which correspond to pairs in the previous generation (e.g., A and
B) and (ii) pairs which do not correspond to pairs in the previous
generation (e.g., D and E). It is obvious that as we move from one
generation to the next, the conductance between pairs of type (i)
increases because there are more paths between the points then in the
previous generation.  On the other hand, the conductance between the
pairs of type (ii) are lower on average than between the pairs present
in the previous generation because on average the shortest path between
the two points is longer than between the pairs in the previous
generation.  However, for any given $r$, the shortest path between any
two points has a fixed upper bound independent of the generation.  Due
to this bound on the shortest path, the net result is that the average
conductivity increases with succeeding generations.  This is shown in
Fig.~\ref{fig3}(d) in which we plot the average conductivity calculated
exactly for generations 1 to 6 for $r=1$, 2 and 4.

\section{Shortest Path Distribution}

In order to understand why the average conductance of the percolation
backbone decreases with increasing $M_B$, we must (i) recognize that the
conductance is strongly correlated with the shortest path between the
two points and (ii) study $P(\ell|M_B,r)$, the distribution of shortest
paths between the two points for a given backbone mass. Hence we next
create the $P(\ell|M_B,r)$ probability distribution, binning the data
logarithmically by taking the average over samples centered at
$\log_2\ell$.

Figure \ref{fig4}(a) shows the results of the simulations for
$P(\ell|M_B,r)$ for $r=1$ for various backbone masses. The plots
collapse, the only difference in the plots being the values of the upper
cut-offs due to the finite size of the backbone. Figure \ref{fig1}
illustrates how the size of the backbone constrains the possible values
of the shortest path. For all values of $M_B$, a section of each plot in
Fig.~\ref{fig4}(a) exhibits power law behavior. In Fig.~\ref{fig4}(b),
we show the distributions $P(\ell|M_B,r)$ for different $r$ and a given
$M_B$. In Fig.~\ref{fig4}(c) we see that when scaled with
$r^{d_{\mbox{\scriptsize min}}}$ the plots collapse, so we can write
$P(\ell|M_B,r)$ in the scaling form
\begin{equation}
\label{f1}
P(\ell|M_B,r)\sim {1\over r^{d_{\mbox{\scriptsize min}}}}\left(
{\ell\over r^{d_{\mbox{\scriptsize min}}}}\right)^{-\psi}.
\end{equation}
An expression for $\psi$ can be found by recognizing that we can write
the well-studied distribution $P(\ell|r)$, the probability that the
shortest path between two points separated by Euclidean distance $r$ is
$\ell$, independent of $M_B$, as
\begin{equation}
\label{f2}
P(\ell|r)=\int_{c_\ell}^\infty P(\ell|M_B,r)P(M_B|r)dM_B,
\end{equation}
where (i) $P(M_B|r)$ is the distribution of backbone masses given
distance $r$ between the points which determine the backbone and (ii)
$c_\ell$ is the lower cutoff on $M_B$ given $\ell$. $P(M_B|r)$ has the
form
\cite{Barthelemy99}
\begin{equation}
\label{f3}
P(M_B|r)\sim {1\over r^{d_B}}\left( {M_B\over r^{d_B}}\right)^{-\tau_B},
\qquad\qquad [r\ll L]
\end{equation}
where $d_B$ is the fractal dimension of the backbone and $\tau_B=d/d_B$
is the exponent for the blob size distribution \cite{Herrmann86}. From
Ref.~\cite{Dokholyan98}
\begin{equation}
\label{f4}
P(\ell|r)\sim {1\over r^{d_{\mbox{\scriptsize min}}}}\left( {\ell\over
r^{d_{\mbox{\scriptsize min}}}}\right)^{-g_\ell},
\end{equation}
where $d_{\mbox{\scriptsize min}}$ is the fractal dimension of the
shortest path. Since $\ell\sim r^{d_{\mbox{\scriptsize min}}}$ and
$M_B\sim r^{d_B}$, implying $\ell\sim M_B^{d_{\mbox{\scriptsize
min}}/d_B}$, the lower cutoff $c_\ell$ in Eq.~(\ref{f2}) scales as
\begin{equation}
\label{f4a}
c_\ell\sim\ell^{d_B/d_{\mbox{\scriptsize min}}}. 
\end{equation}
As $L\to\infty$, the upper cutoff is $\infty$ because the maximum
backbone mass is not constrained by the length of the shortest
path. Substituting Eqs.~(\ref{f3}), (\ref{f4}), and (\ref{f4a}) into
Eq.~(\ref{f2}), and equating powers of $r$ (or powers of $\ell$) of the
left and right hand sides of the resulting equation, we find
\begin{equation}
\label{f6}
\psi=g_\ell-{d_B\over d_{\mbox{\scriptsize min}}}(\tau_B-1).
\end{equation}
Using $\tau_B=d/d_B$ and the values $g_\ell=2.04$ \cite{Dokholyan98} and
$d_{\mbox{\scriptsize min}}=1.13$ \cite{GrassbergerXX}, we find
$\psi=1.72$, in good agreement with our simulation result
\begin{equation}
\label{f6a}
\psi=1.7\pm 0.05.
\end{equation}

\section{Average Conductance}

\begin{figure}

\centerline{ \epsfxsize=7.0cm
\rotate[r]{ \epsfbox{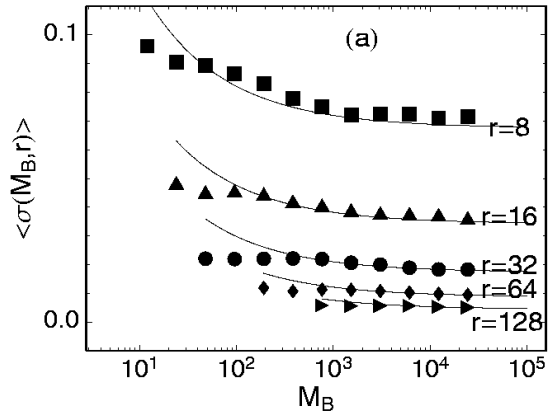} }
\vspace*{1.0cm}
}
\centerline{ \epsfxsize=7.0cm
\rotate[r]{ \epsfbox{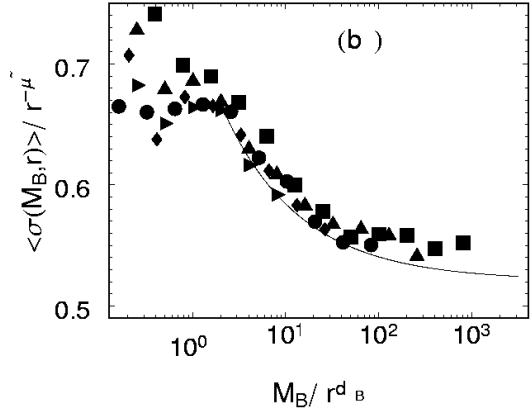} }
\vspace*{1.0cm}
}
\centerline{ \epsfxsize=7.0cm
\rotate[r]{ \epsfbox{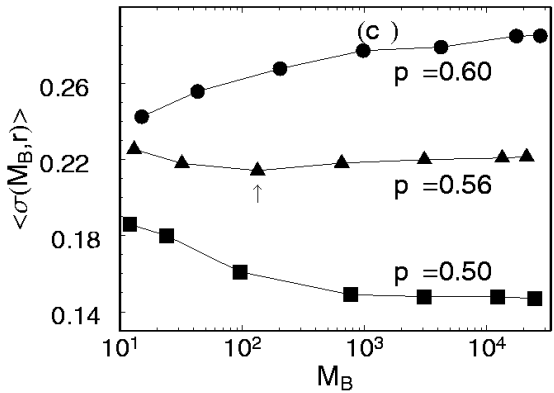} }
\vspace*{1.0cm}
}

\caption{
Average conductance versus backbone mass.  (a) Simulation results at the
percolation threshold for $r=8$, 16, 32, 64, and 128, where $r$ is the
distance between the two sites $A$ and $B$; the adjacent lines are the
theoretical results. For large $r$, the curves for the simulation
results and the corresponding curves for the theoretical results
coincide for large $M_B$. (b) Plots of scaled backbone conductance for
$r=8$, 16, 32, 64, and 128, scaled in accordance with
Eqs.~(\protect\ref{f12}) and (\protect\ref{f13}). The solid line is a
plot of Eq.~(\protect\ref{f12}) with parameters $a$ and $b$ chosen as
0.9 and 6 respectively to best fit the values for $r=128$
(right-pointing triangles). The collapse to this line for the lower
values of $r$ improves with increasing $r$. (c) Average conductance for
$r=4$ for $p=0.50$, 0.56, and 0.60. For $p=0.56$, the conductance as a
function of $M_B$ is not monotonic but rather has a minimum indicated by
the arrow.}
\label{fig2}
\end{figure}

\begin{figure}
\centerline{ \epsfxsize=14.0cm
\rotate[r]{ \epsfbox{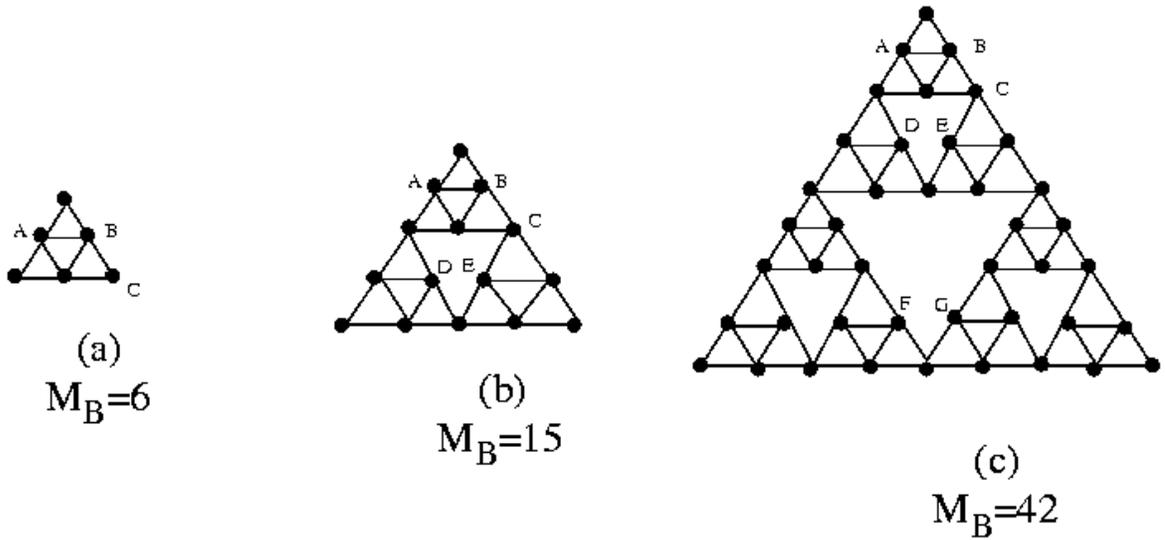} }
\vspace*{1.0cm}
}
\begin{multicols}{2}

\centerline{ \epsfxsize=7.0cm
\rotate[r]{ \epsfbox{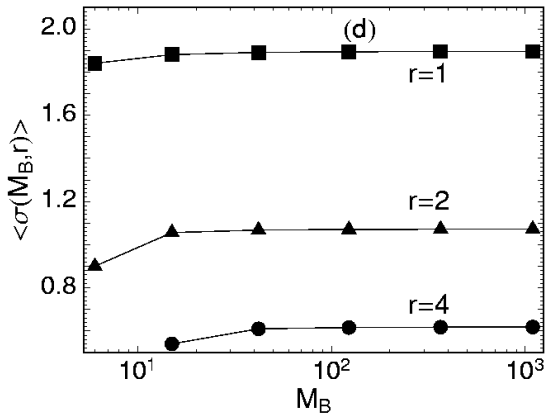} }
\vspace*{1.0cm}
}

\caption{
(a)--(c) Three generations of the Sierpinski gasket.  At each successive
generation, there are two types of pairs of points: pairs which
correspond to pairs in the previous generation and pairs which do not.
For example, in the second generation the pairs AB and AC are present in
the previous generation but the pair DE has no corresponding pair in the
previous generation.  Similarly, the pairs AB, AC, and DE in the third
generation correspond to pairs in the second generation, but the pair FG
does not.  Because all points are multiply connected, the mass of the
backbone between any two points in each generation is equal to the mass
of the entire gasket. (d) Average conductance between all pairs of
points separated by distance $r$ on the Sierpinski gasket versus the
gasket mass.  The points correspond to successive generations of the
Sierpinski gasket.  }
\label{fig3}

\end{multicols}
\end{figure}

\begin{figure}
\centerline{ \epsfxsize=7.0cm
\rotate[r]{ \epsfbox{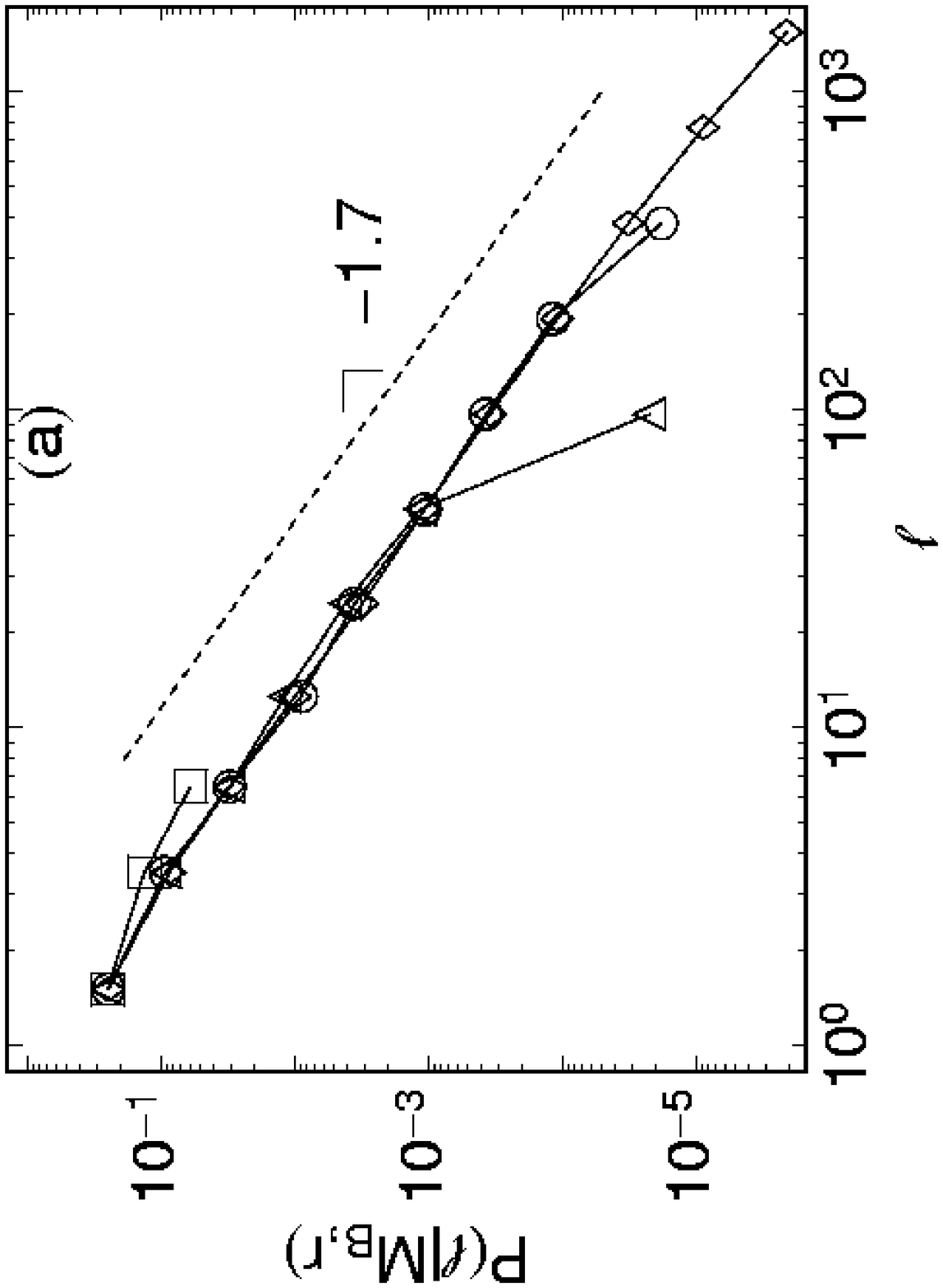} }
\vspace*{1.0cm}
}
\centerline{ \epsfxsize=7.0cm
\rotate[r]{ \epsfbox{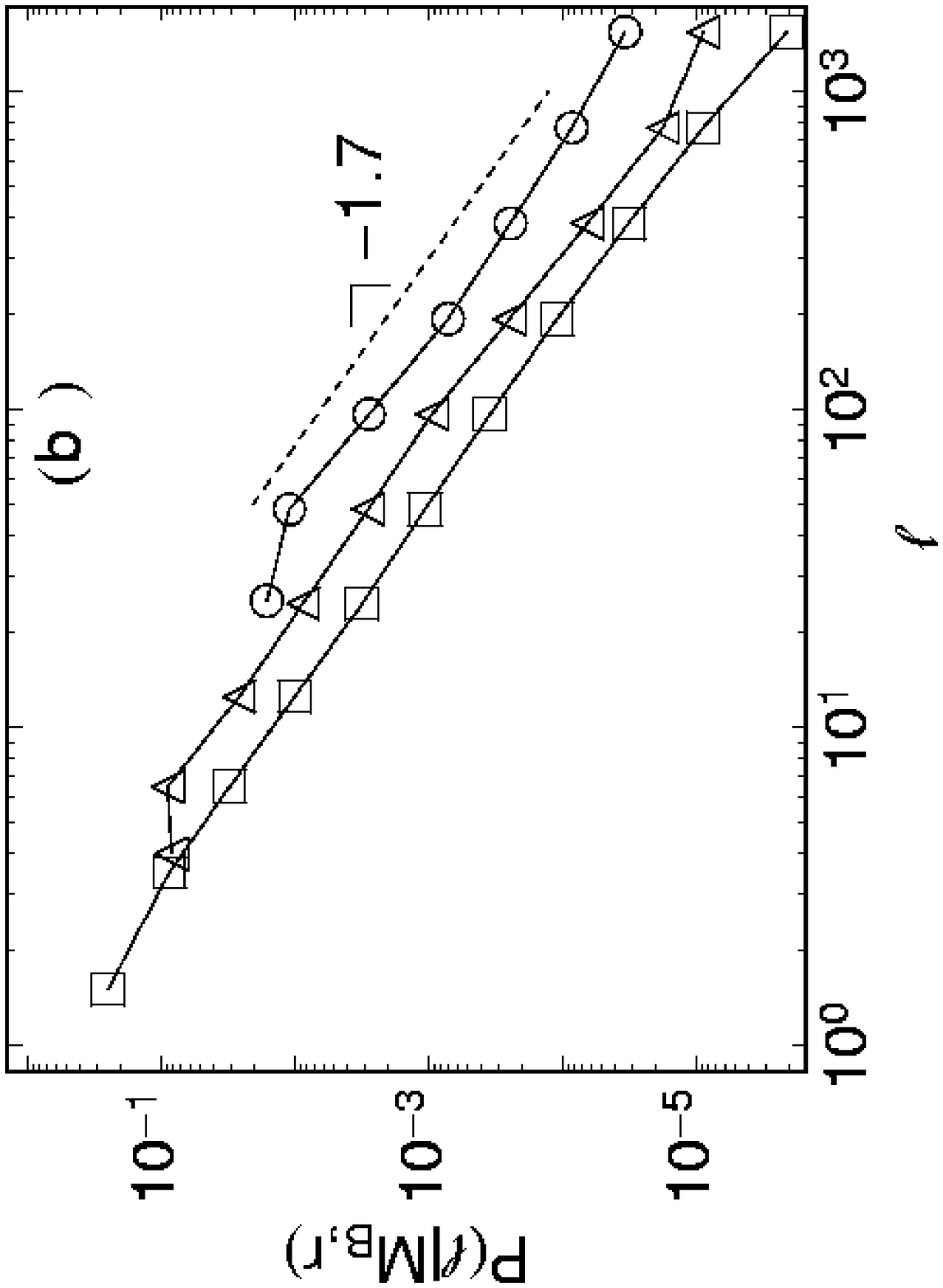} }
\vspace*{1.0cm}
}
\centerline{ \epsfxsize=7.0cm
\rotate[r]{ \epsfbox{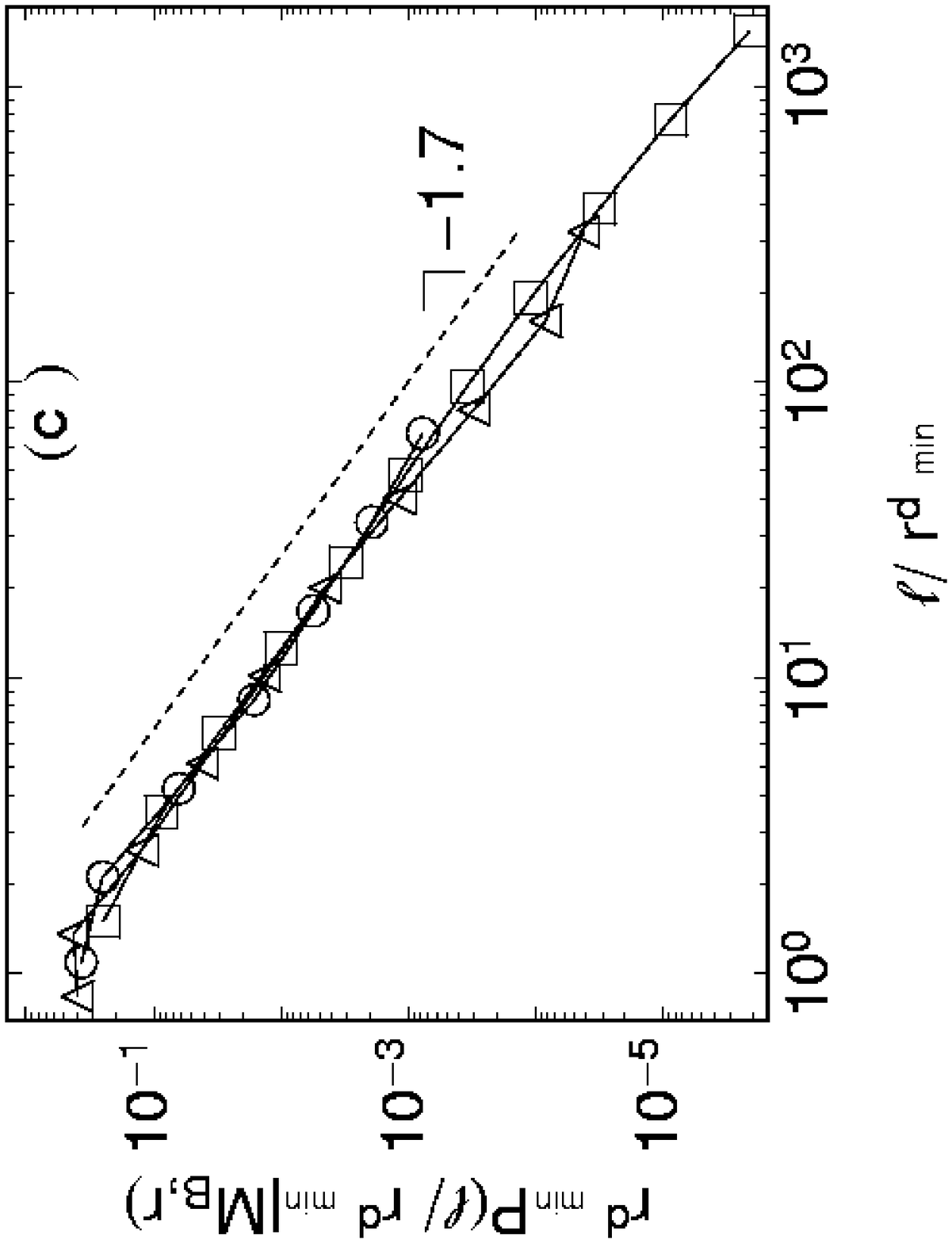} }
\vspace*{1.0cm}
}

\caption{
Distribution of shortest paths between $A$ and $B$. (a) $P(\ell|M_B,r)$,
the probability that the length of the shortest path between two points
separated by distance $r$ is $\ell$ for a given backbone mass, $M_B$.
All plots are for $r=1$, for values of $M_B$ of 6 (squares), 96
(diamonds), 1536 (circles), and 24576 (triangles). The plots for the
various $M_B$ differ by the points at which they cut off.  (b)
$P(\ell|M_B,r)$ for $r=1$ (squares), 4 (diamonds), and 16 (circles) for
a single backbone size of 24576. (c) When scaled by
$r^{d_{\mbox{\scriptsize min}}}$, the plots collapse. The dashed line is
constructed to have a slope of $-1.7$; see Eq.~(\protect\ref{f6a}).}
\label{fig4} 
\end{figure}

We can now calculate the average conductance.  Since $\sigma$ is
strongly correlated with $\ell$, and since $\sigma$ scales with $r$ as
$r^{-\tilde\mu}$ and $\ell$ scales with $r$ as $r^{d_{\mbox{\scriptsize
min}}}$, we have 
\begin{equation}
\label{f6b}
\langle\sigma\rangle\sim\ell^{-\tilde\mu/d_{\mbox{\scriptsize
min}}}. 
\end{equation}
Then using the fact that $P(\ell|M_B,r)\sim\ell^{-\psi}$ we have
\begin{eqnarray}
\label{f7}
\nonumber P(\sigma|M_B,r) &\sim& P(\ell=\sigma^{-d_{\mbox{\scriptsize
min}}/\tilde\mu}|M_B,r){d\ell\over d\sigma} \\
                &\sim& \sigma^{(\psi-1)(d_{\mbox{\scriptsize
min}}/\tilde\mu)-1}=\sigma^z,
\end{eqnarray}
where 
\begin{equation}
\label{f7a}
z\equiv (\psi-1)(d_{\mbox{\scriptsize min}}/\tilde\mu)-1=-0.17.
\end{equation}
Now $P(\ell|M_B,r)$ is nonzero only for 
\begin{equation}
\label{f7b}
(ar)^{d_{\mbox{\scriptsize min}}}\buildrel < \over\sim\ell\buildrel <
\over\sim (bM_B)^{d_{\mbox{\scriptsize min}}/d_B},
\end{equation}
where $a$ and $b$ are constants. Hence using
$\langle\sigma\rangle\sim\ell^{-\tilde\mu/d_{\mbox{\scriptsize min}}}$,
we find $P(\sigma|M_B,r)$ is nonzero for
\begin{equation}
\label{f7c}
(bM_B)^{(d_{\mbox{\scriptsize
min}}/d_B)(-\tilde\mu/d_{\mbox{\scriptsize min}})}=
(bM_B)^{-\tilde\mu/d_B}\buildrel < \over\sim\sigma\buildrel < \over\sim
(ar)^{-\tilde\mu}.
\end{equation}
Using these bounds to normalize the distribution, we find
\begin{equation}
\label{f8}
P(\sigma|M_B,r)={(z+1)\sigma^z\over
(ar)^{-\tilde\mu(z+1)}-(bM_B)^{(-\tilde\mu/d_B)(z+1)}}.
\end{equation}
Then
\begin{eqnarray}
\label{f9}
\nonumber\langle\sigma(M_B,r)\rangle &=& \int_{(bM)^{-\tilde\mu/d_B}}^{(ar)^{-\tilde\mu}}\sigma
P(\sigma|M_B,r)d\sigma \\ \nonumber\\
                                     &=& {z+1\over z+2}(ar)^{-\tilde\mu}
{1-\left[{(bM_B)^{1/d_B}\over
ar}\right]^{-\tilde\mu(z+2)}\over 1-\left[{(bM_B)^{1/d_B}\over
ar}\right]^{-\tilde\mu(z+1)}}.
\end{eqnarray}
Thus as $M_B$ goes to infinity, $\langle\sigma(M_B,r)\rangle$ decreases
asymptotically to a constant as
\begin{equation}
\label{f11}
\langle\sigma(M_B,r)\rangle\sim {z+1\over z+2}(ar)^{-\tilde\mu}
\left[1+\left[{(bM_B)^{1/d_B}\over
ar}\right]^{-\tilde\mu(z+1)}\right].
\end{equation}
By considering the asymptotic dependence of
$\langle\sigma(M_B,r)\rangle$ on $M_B$, we can reasonably fit the
simulation results by choosing the parameters $a$ and $b$ in
Eq.~(\ref{f9}) to be 0.9 and 6, respectively. Using these values for $a$
and $b$, in Fig.~\ref{fig2}(a) we plot $\langle\sigma\rangle$ from
Eq.~(\ref{f9}) for multiple values of $r$ and find that agreement with
the simulation results improves with increasing $r$. For large $r$, the
curves for the simulation results and the curves for the theoretical
results are coincident at large $M_B$. The poor results for small $r$
are due to corrections-to-scaling not included in our derivation (e.g.,
for small $r$, there are significant corrections-to-scaling for the
relations $\sigma\sim r^{-\tilde\mu}$ and $M_B\sim r^{d_B}$
\cite{Grassberger99}).

Equation (\ref{f9}) can be re-cast in terms of the scaled variable
$x\equiv M_B/r^{d_B}$ as
\begin{equation}
\label{f12}
\sigma(x,r)={z+1\over z+2}(ar)^{-\tilde\mu}f(x),
\end{equation}
where
\begin{equation}
\label{f13}
f(x)={1-\left({b\over a^{d_B}}x\right)^{-(\tilde\mu/d_B)(z+2)}\over
1-\left({b\over a^{d_B}}x\right)^{-(\tilde\mu/d_B)(z+1)}}.
\end{equation}
In Fig.~\ref{fig2}(b), we plot the average conductance scaled in
accordance with Eqs.~(\ref{f12}) and (\ref{f13}). The expected collapse
improves with increasing $r$ for the same reason as noted above.

Above the percolation threshold, for backbones of size larger than the
correlation length, the strong correlation between the conductance and
the shortest path breaks down and we expect the conductance to {\it
increase\/} with the mass of the backbone, as is the case in non-random
systems. This is seen in Fig.~\ref{fig2}(c), where we plot conductance
versus backbone mass for the bond occupation probabilities $p=0.56$ and
$p=0.60$, which are above the percolation threshold and, for comparison,
conductance {\it at\/} the percolation threshold, $p=0.50$ \cite{text1}.
Figure \ref{fig2}(c) shows that for $p=0.60$, all backbone masses
sampled are of size greater than the correlation length and the
conductance increases monotonically. For $p=0.56$, the smaller backbone
masses are of size less than the correlation length and
Fig.~\ref{fig2}(c) shows that the conductance initially decreases; for
larger backbone masses, however, the sizes of the backbones are greater
than the correlation length and Fig.~\ref{fig2}(c) shows that the
conductance then increases.

\section{Discussion}

The derivation of Eq.~(\ref{f9}) and its agreement with the results of
our simulations confirm our understanding of why the average conductance
decreases with increasing backbone mass: the smaller contributions to
the average conductance from the longer minimal paths possible in the
clusters with larger backbone size cause the average conductance to be
smaller. Our derivation was not specific to two dimensions, and should
also hold in higher dimensions.

\section*{Acknowledgments}

We thank J. Andrade for helpful discussions, and BP Amoco for financial
support.

\newpage

%\widetext

\end{multicols}

\newpage
\begin{multicols}{2}

\end{multicols}

\newpage

\newpage
\begin{multicols}{2}

\end{multicols}


\begin{references}
\bibitem{text1}
Because, above the percolation threshold, the backbone mass is strongly
correlated with the system size, we create backbones of a given mass by
varying the system size. The simulations of Fig.~2(c) for the plots
above the percolation were obtained choosing $L=6$, 10, 20, 40, 80, 160,
and 200.
\end{references}
\end{document}